\PassOptionsToPackage{unicode}{hyperref}
\PassOptionsToPackage{hyphens}{url}
\documentclass[
  11pt,
]{article}
\usepackage{xcolor}
\usepackage[margin=1in]{geometry}
\usepackage{amsmath,amssymb}
\usepackage{iftex}
\ifPDFTeX
  \usepackage[T1]{fontenc}
  \usepackage[utf8]{inputenc}
  \usepackage{textcomp} % provide euro and other symbols
\else % if luatex or xetex
  \usepackage{unicode-math} % this also loads fontspec
  \defaultfontfeatures{Scale=MatchLowercase}
  \defaultfontfeatures[\rmfamily]{Ligatures=TeX,Scale=1}
\fi
\usepackage{lmodern}
\ifPDFTeX\else
\fi
\IfFileExists{upquote.sty}{\usepackage{upquote}}{}
\IfFileExists{microtype.sty}{% use microtype if available
  \usepackage[]{microtype}
  \UseMicrotypeSet[protrusion]{basicmath} % disable protrusion for tt fonts
}{}
\makeatletter
\@ifundefined{KOMAClassName}{% if non-KOMA class
  \IfFileExists{parskip.sty}{%
    \usepackage{parskip}
  }{% else
    \setlength{\parindent}{0pt}
    \setlength{\parskip}{6pt plus 2pt minus 1pt}}
}{% if KOMA class
  \KOMAoptions{parskip=half}}
\makeatother
\usepackage{longtable,booktabs,array}
\usepackage{caption}
\usepackage{calc} % for calculating minipage widths
\usepackage{etoolbox}
\makeatletter
\patchcmd\longtable{\par}{\if@noskipsec\mbox{}\fi\par}{}{}
\makeatother
\IfFileExists{footnotehyper.sty}{\usepackage{footnotehyper}}{\usepackage{footnote}}
\makesavenoteenv{longtable}
\usepackage{graphicx}
\makeatletter
\newsavebox\pandoc@box
\newcommand*\pandocbounded[1]{% scales image to fit in text height/width
  \sbox\pandoc@box{#1}%
  \Gscale@div\@tempa{\textheight}{\dimexpr\ht\pandoc@box+\dp\pandoc@box\relax}%
  \Gscale@div\@tempb{\linewidth}{\wd\pandoc@box}%
  \ifdim\@tempb\p@<\@tempa\p@\let\@tempa\@tempb\fi% select the smaller of both
  \ifdim\@tempa\p@<\p@\scalebox{\@tempa}{\usebox\pandoc@box}%
  \else\usebox{\pandoc@box}%
  \fi%
}
\def\fps@figure{htbp}
\makeatother
\providecommand{\tightlist}{%
  \setlength{\itemsep}{0pt}\setlength{\parskip}{0pt}}
\usepackage{amssymb}
\usepackage{amsmath}
\usepackage{fontspec}
\usepackage{seqsplit}
\usepackage[htt]{hyphenat}
\usepackage{xurl}
\usepackage{graphicx}
\setkeys{Gin}{width=0.85\linewidth,keepaspectratio}
\usepackage{float}
\usepackage{caption}
\usepackage{newunicodechar}
\newunicodechar{∈}{\ensuremath{\in}}
\newunicodechar{∞}{\ensuremath{\infty}}
\newunicodechar{≤}{\ensuremath{\leq}}
\newunicodechar{≥}{\ensuremath{\geq}}
\newunicodechar{≠}{\ensuremath{\neq}}
\newunicodechar{≈}{\ensuremath{\approx}}
\newunicodechar{×}{\ensuremath{\times}}
\newunicodechar{−}{\ensuremath{-}}
\newunicodechar{·}{\ensuremath{\cdot}}
\newunicodechar{±}{\ensuremath{\pm}}
\newunicodechar{α}{\ensuremath{\alpha}}
\newunicodechar{β}{\ensuremath{\beta}}
\newunicodechar{σ}{\ensuremath{\sigma}}
\newunicodechar{Σ}{\ensuremath{\Sigma}}
\newunicodechar{π}{\ensuremath{\pi}}
\newunicodechar{→}{\ensuremath{\rightarrow}}
\newunicodechar{⁻}{\ensuremath{{}^{-}}}
\newunicodechar{⁰}{\ensuremath{{}^{0}}}
\newunicodechar{⁸}{\ensuremath{{}^{8}}}
\newunicodechar{₀}{\ensuremath{{}_{0}}}
\newunicodechar{₁}{\ensuremath{{}_{1}}}
\newunicodechar{₂}{\ensuremath{{}_{2}}}
\newunicodechar{₃}{\ensuremath{{}_{3}}}
\newunicodechar{₄}{\ensuremath{{}_{4}}}
\newunicodechar{₅}{\ensuremath{{}_{5}}}
\newunicodechar{₆}{\ensuremath{{}_{6}}}
\newunicodechar{₇}{\ensuremath{{}_{7}}}
\newunicodechar{₈}{\ensuremath{{}_{8}}}
\newunicodechar{₉}{\ensuremath{{}_{9}}}
\usepackage{bookmark}
\IfFileExists{xurl.sty}{\usepackage{xurl}}{} % add URL line breaks if available
\makeatletter
\@ifundefined{xmpquote}{}{}
\makeatother
\hypersetup{
  hidelinks,
  pdfcreator={LaTeX via pandoc}}

\author{}
\date{}

\begin{document}

\section{Auditing Collector-Generated Graduation Labels on Pump.fun:
Measurement Error and Temporal
Non-Generalization}\label{auditing-collector-generated-graduation-labels-on-pump.fun-measurement-error-and-temporal-non-generalization}

\textbf{Author:} Arati Uday Kamat \textbf{ORCID:} 0009-0000-4781-312X
\textbf{Affiliation:} Independent researcher, United States
\textbf{Pre-registration:}
\texttt{PREREGISTRATION\_v2.3\_FROZEN\_2026-08-20.md}, SHA-256
\texttt{ee1702b3510d21c51c8f4ab95660e7804f4a93ab09e21aacb3a70b348a90b40a}
\textbf{Frozen inputs:} \texttt{red\_pump\_2026\_v1\_launches.jsonl.gz}
and \texttt{red\_pump\_2026\_v1\_outcomes.csv.gz} (SHA-256s listed in
§11). \textbf{Manuscript version:} 5th draft (corrected). ``Manuscript
v5'' and the reproducibility ``package v1.5'' refer to different version
tracks (manuscript vs data-and-code archive). \textbf{Data
availability:} Frozen inputs, code, and reproducibility package are
archived at Zenodo, concept DOI
\href{https://doi.org/10.5281/zenodo.20633486}{10.5281/zenodo.20633486}.

\subsection{Structured abstract}\label{structured-abstract}

\textbf{Purpose.} To examine whether collector-generated terminal labels
on the pump.fun platform can be interpreted as platform-side graduation
outcomes, and whether a pre-registered logistic association model
generalises across a subsequent temporal holdout of the same cohort.

\textbf{Design/methodology/approach.} A frozen primary cohort of 749,816
unique mints recorded by a V3 off-chain collector between 12 May 2026
and 10 June 2026 is analysed. The design combines (i) a source-code and
reconstructed-state measurement audit of the collector's
terminal-classification mechanism with (ii) a pre-registered locked
logistic association model fitted on a 15-day development cohort and
evaluated on a subsequent 14-day validation cohort under the
pre-registered stability-gate framework. The deposited stability-gate
evaluation emitted nine automated stability-gate assessments for Gates
B, C, D, and G. Calendar-day cluster-robust covariance and day-block
bootstrap procedures assess uncertainty.

\textbf{Findings.} Development discrimination is high (AUROC 0.8594),
but the model does not generalise to the held-out validation cohort
(AUROC 0.4642; 500 of 500 successful bootstrap replicates; 95\%
percentile interval {[}0.4112, 0.5196{]} containing 0.5000). Validation
calibration deteriorates substantially (slope 0.013, intercept +2.816);
the market-cap functional form is not stable across pre-registered
specifications; and of the nine automated evaluations, two pass, six
fail, and one is not evaluable. The collector's TIMEOUT label does not
establish platform-side non-graduation.

\textbf{Originality/value.} The paper treats outcome ascertainment as a
first-order measurement problem rather than assuming that an off-chain
collector's terminal classification measures the platform-side event. It
pairs this audit with a pre-registered held-out temporal-generalisation
test and reports the resulting negative finding as-registered, alongside
a fully reproducible correction history.

\textbf{Keywords.} crypto microstructure, temporal validation,
measurement error, pre-registration, calibration failure, Solana

\textbf{JEL classification.} G14 (Information and Market Efficiency;
Event Studies); C52 (Model Evaluation, Validation, and Selection); C55
(Modeling with Large Data Sets)

\subsection{1. Introduction}\label{introduction}

Pump.fun is a Solana-based memecoin issuance platform where each newly
minted token accretes a bonding-curve market-capitalization (mcap)
balance that, on reaching a platform-side threshold (``graduation''),
transitions to an automated market-maker pool. Community-recorded
metrics of that pool transition are typically compiled from a top-50
feed reconstructed by an off-chain collector, the same feed used by this
study's frozen input (see §2, §5). Two methodological facts about that
literature have been under-discussed.

\textbf{Fact 1, the ``outcome'' is not observable directly.} Analysts do
not receive per-mint bonding-curve completion events from the platform.
They receive whatever their off-chain collector chose to record, a
design choice of the collector, not of the platform. This dataset's V3
collector polls the platform's top-50 newest-coins feed on a 60-second
loop and reconstructs terminal states from what appears in successive
poll responses (The V3 source-code audit is archived at
\texttt{04\_frozen\_inputs/PHASE1\_RECONSTRUCTION\_2026-08-20/preservation/}).
The observation-window and top-50 resurfacing limits of this mechanism
are documented in
\texttt{04\_frozen\_inputs/OUTCOME\_ASCERTAINMENT\_MECHANISM.md}. A
collector recording that a mint ``graduated'' is stating that, at some
poll, it observed that mint with \texttt{complete\ ===\ true}. A
collector recording ``timeout'' is stating that no such observation
occurred within the collector's own 24-hour wall-clock timeout.
\textbf{The recorded terminal label is a joint function of the
platform-side event AND the collector's polling design.}

\textbf{Fact 2, associations estimated within one temporal slice need
not extend.} The general point that in-sample discrimination on a single
slice can misrepresent behaviour on subsequent data is well-documented
in the clinical-prediction and calibration literature (Steyerberg, 2019;
Harrell, 2015; Van Calster et al., 2019). Without a pre-registered
held-out temporal-generalization test (Nosek et al., 2018), a
within-slice model that discriminates well can still misdirect readers
about behaviour beyond its fit window.

\textbf{Gap.} In the pump.fun setting specifically, no comparable audit
was identified among the sources reviewed for this revision. Those
sources included Solana Foundation documentation (Solana Foundation,
n.d.) and this study's frozen record of the pump.fun top-50 endpoint
response contract reconstructed from the V3 collector's polling logs.
None of the reviewed sources audits the collector-generated outcome
label against the collector's own source code while also pre-registering
and reporting a locked held-out temporal-generalization test. This paper
does not claim exhaustive literature coverage; a systematic search of
the peer-reviewed pump.fun-specific measurement-audit literature was not
conducted for this revision, so this gap statement is intentionally
scoped to the sources reviewed.

\textbf{Contribution.} Two pre-registered analyses are conducted on a
frozen pump.fun launch corpus (2026-05-12 through 2026-06-10):

\begin{itemize}
\tightlist
\item
  \textbf{Aim 1 (measurement audit).} The V3 collector's source code and
  reconstructable state are audited
  (\texttt{04\_frozen\_inputs/PHASE1\_RECONSTRUCTION\_2026-08-20/}) to
  quantify where the collector-recorded terminal label can and cannot
  correspond to platform-side graduation, and to characterize the
  observation-window and top-50 resurfacing limitations.
\item
  \textbf{Aim 2 (temporal-generalization test).} The v2.3 pre-registered
  logistic association model is fit on a 15-day development cohort,
  frozen, and evaluated on a subsequent 14-day validation cohort under
  the pre-registered stability-gate framework (Gates B, C, D, G
  evaluated automatically as nine tally rows; Gates A, E, H addressed
  narratively in §7). The result is reported as-registered, whether it
  passes or fails.
\end{itemize}

\textbf{Why v5 replaces v1.3.} The prior manuscript version (Kamat,
2026; v1.3, Zenodo DOI
\href{https://doi.org/10.5281/zenodo.21908375}{10.5281/zenodo.21908375})
reported a 24-hour graduation rate, a ``3.18× decline'' between two
temporal slices, wallet-intent narratives around seed-buys, and
Kaplan-Meier / Cox specifications on collector-reconstructed censoring
times. Every one of those claims is retracted here because each depends
on interpreting the collector-recorded terminal classification as the
platform-side event, on reconstructed censoring times that violated the
pre-registered adequacy ceiling, or on a wallet-intent variable the
design does not observe. This v5 replaces v1.3 in full; the correction
history is documented in §12 and Appendix A of the deposited artefact.

The scientific contribution of v5 is deliberately narrower and
negative-result-shaped: a documented collector-terminal-label
measurement audit and a reproducible failure of prespecified held-out
temporal generalization. Both findings are reportable under the frozen
v2.3 pre-registration with the §14 language-restriction gate satisfied.

\subsection{2. Data}\label{data}

\subsubsection{2.1 Frozen input corpus}\label{frozen-input-corpus}

Two files, sealed under SHA-256 before any modelling:

\begin{itemize}
\tightlist
\item
  \texttt{red\_pump\_2026\_v1\_launches.jsonl.gz} (SHA-256
  \texttt{04294037\hspace{0pt}9e8c897a\hspace{0pt}c97403e6\hspace{0pt}b25a5b30\hspace{0pt}2fb32b69\hspace{0pt}02a8fc0c\hspace{0pt}ef4ab70a\hspace{0pt}c11e8f84}):
  one line per collector-observed mint launch event, with
  first-detection timestamp t, \texttt{has\_twitter},
  \texttt{has\_website}, \texttt{has\_telegram} flags,
  \texttt{initial\_market\_cap\_sol}, \texttt{description\_length},
  \texttt{created\_timestamp}, and provenance fields.
\item
  \texttt{red\_pump\_2026\_v1\_outcomes.csv.gz} (SHA-256
  \texttt{c0a327ea\hspace{0pt}442d91c6\hspace{0pt}f970b2ba\hspace{0pt}d9a2a9b7\hspace{0pt}78e163d8\hspace{0pt}c3eb38f7\hspace{0pt}1eccd3e9\hspace{0pt}2209a974}):
  one row per collector-recorded outcome, carrying the terminal label
  (\texttt{GRADUATED} or \texttt{TIMEOUT}), \texttt{graduated\_at\_ms},
  and the deduplication and malformed-row flags established in the
  frozen data schema.
\end{itemize}

Cohort boundaries are locked at the V3 collector-version boundary:
primary V3-only period {[}2026-05-12T13:49:06Z, 2026-06-10T18:11:08Z{]}.
Cohorts preceding V3 (V1/V2) are retained as descriptive-only
sensitivities.

\subsubsection{2.2 Cohort flow}\label{cohort-flow}

The deterministic ledger \texttt{cohort\_flow.csv} records each unique
mint's \texttt{is\_V3\_primary\_window}, \texttt{dedup\_action},
\texttt{has\_valid\_terminal\_row}, \texttt{outcome\_label},
\texttt{enters\_primary}, and exclusion reason per pre-reg §3.1. After
the four conjunctive inclusion conditions:

n\_primary = 749,816 mints, n\_GRADUATED = 1,597, n\_TIMEOUT = 748,219.

Development / validation split at the pre-registered boundary t =
2026-05-27T00:00:00Z:

n\_dev = 387,152 (51.6\%), n\_val = 362,664 (48.4\%).

n\_GRADUATED,dev = 794 (prevalence0.205\%), n\_GRADUATED,val = 803
(prevalence0.221\%).

The two-week temporal boundary places roughly half the cohort in each
split with near-identical marginal prevalence, a design choice made in
the pre-registration before validation-set metrics were computed.

\begin{figure}
\centering
\pandocbounded{\includegraphics[keepaspectratio]{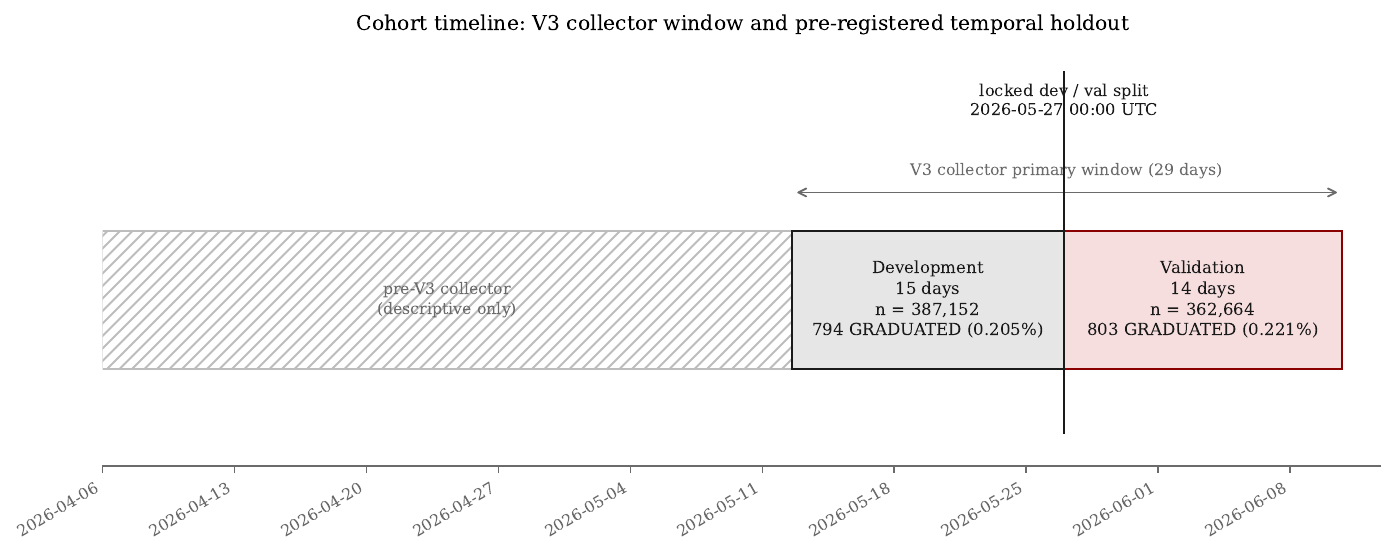}}
\caption{Cohort timeline. Pre-V3 window (hatched) is excluded; V3
primary window (2026-05-12 to 2026-06-10) is split at the pre-registered
boundary 2026-05-27 into development (15 days, 387,152 mints) and
validation (14 days, 362,664 mints) cohorts.}
\end{figure}

\subsubsection{2.3 Outcome ascertainment}\label{outcome-ascertainment}

Per the pre-registered §2 disclaimer (reproduced verbatim from the
frozen pre-registration):

\begin{quote}
A mint whose bonding curve fills after scrolling off the top-50 feed
will be recorded as GRADUATED only if it subsequently appears in a
polled top-50 response with \texttt{complete\ ===\ true} before its
24-hour timeout is processed. Otherwise, the collector records TIMEOUT.
Consequently, TIMEOUT does not establish that platform-side graduation
did not occur. Association estimates below describe association with the
collector-recorded terminal classification within the V3 collector
cohort, not association with true 24-hour graduation on pump.fun.
\end{quote}

The endpoint Y\_i ∈ \{0,1\} is coded Y\_i = 1 if
\texttt{outcome\ ==\ "GRADUATED"} and Y\_i = 0 if
\texttt{outcome\ ==\ "TIMEOUT"}. This is the collector's deterministic
terminal classification. The mechanism by which it can and cannot
correspond to platform-side graduation is documented in the deposited
outcome-ascertainment specification (deposited with the artefact) and
revisited in §5.

\subsection{3. Endpoint definition}\label{endpoint-definition}

The primary endpoint is the locked binary Y\_i per §2.3. Reported
analyses use Y\_i directly. The following alternative endpoints, present
in v1.3, are \textbf{not} used here:

\begin{itemize}
\tightlist
\item
  ``24-hour graduation rate'' (v1.3): treated collector-recorded
  observation as the platform-side event; withdrawn.
\item
  Time-to-graduation with reconstructed censoring (v2.2): the
  reconstruction-adequacy stopping rule triggered at 2.6925\%
  (pre-registered ceiling 2\%); withdrawn from primary and retained only
  as a supplementary measurement audit.
\item
  Cox proportional-hazards specification (v1.3): withdrawn for the same
  reconstruction-symmetry violation.
\end{itemize}

The v2.3 pre-registration §1.2 supersedes the v2.2 discrete-time hazard
model with binary logistic regression on the terminal label. All
downstream metrics are binary-classification metrics.

\subsection{4. Methods}\label{methods}

\subsubsection{4.1 Primary specification (P1,
locked)}\label{primary-specification-p1-locked}

Binary logistic regression, one covariate row per mint, one intercept,
no penalization, no class weighting, no resampling. The design formula
uses the \texttt{patsy} formula grammar (Smith, n.d.), transcribed
verbatim from the frozen pre-registration §5:

\begin{verbatim}
Y ~ has_twitter
  + has_website
  + has_telegram
  + C(mcap_category, Treatment(reference="[29.9999995, 30.0000005)"))
  + log1p(description_length)
  + patsy.bs(day_index, degree=3, include_intercept=False,
              knots=<dev cohort 5/27.5/50/72.5/95 percentiles>,
              lower_bound=day_index_at_2026-05-12T13:49:06Z,
              upper_bound=day_index_at_2026-06-10T18:11:08Z)
  + C(hour, Treatment(reference=0))
  + log1p(launch_intensity_prior_5min)
\end{verbatim}

Reported odds ratios for the P2 spline and P3 log1p specifications at a
given contrast pair are computed as
\texttt{exp(logit(p\_hi)\ -\ logit(p\_lo))} at the two mcap values
holding all other covariates at the development-cohort mean; the full
contrast operator and the deposited \texttt{mcap\_contrasts.csv} schema
are documented in the reproducibility package. The four-part
\texttt{mcap\_category} boundaries are \{(-∞, 29.9999995),
{[}29.9999995, 30.0000005), {[}30.0000005, 31.04), {[}31.04, ∞)\}; the
second (locked reference) captures the very narrow platform-side
graduation threshold neighborhood. The launch-intensity control is
trailing, counted in the half-open interval {[}t\_i - 5 min, t\_i) and
then log(1+x)-transformed. Implementation verification for the
trailing-intensity variable is included in the reproducibility package.

\subsubsection{4.2 Estimation}\label{estimation}

\texttt{statsmodels.GLM(family=Binomial(link=Logit()))} (Seabold and
Perktold, 2010) with \texttt{method="IRLS"}, \texttt{maxiter=200},
\texttt{tol=1e-8}, \texttt{cov\_type="cluster"},
\texttt{cov\_kwds=\{"groups":\ date\_utc,\ "use\_correction":\ True\}}
(cluster-robust covariance per Cameron and Miller, 2015);
\texttt{use\_t=False}. Package versions are locked in
\texttt{environment/requirements.txt} (numpy 2.5.1, pandas 2.3.3, scipy
1.18.0, statsmodels 0.14.6, patsy 1.0.2, scikit-learn 1.9.0).
Development-set \texttt{design\_info} is preserved after fit and reused
unchanged on validation; validation covariate values outside development
bounds are clipped per pre-reg §5.1 and their counts reported.

\subsubsection{4.3 Uncertainty (day-block
bootstrap)}\label{uncertainty-day-block-bootstrap}

Base seed 20260820. Cluster unit is calendar day (\texttt{date\_utc}).
Days are sampled with replacement per the day-block bootstrap (Efron and
Tibshirani, 1993); duplicated sampled days receive new bootstrap cluster
identifiers so cluster-robust covariance treats each replicate-day as an
independent cluster. Development replicates refit the locked P1
specification on a resampled day-block and evaluate the refitted model
on the fixed original development sample; under this fixed-sample
construction the point estimate may lie above the upper percentile of
the interval without indicating a mathematical inconsistency. Validation
replicates apply the locked development-fitted model without refitting
and resample validation-cohort calendar days independently. Percentile
intervals are reported.

Per-replicate seeds derive from a two-level
\texttt{numpy.random.SeedSequence} spawn:

\begin{verbatim}
master = numpy.random.SeedSequence(20260820)
development_seed, validation_seed = master.spawn(2)
development_rngs = [numpy.random.default_rng(s) for s in development_seed.spawn(500)]
validation_rngs  = [numpy.random.default_rng(s) for s in validation_seed.spawn(500)]
\end{verbatim}

Failed replicates are neither replaced nor rerolled. Denominators are
reported as \texttt{successful\ /\ requested}.

\textbf{Bootstrap variance caveat.} The development interval is computed
under a fixed-sample construction (each successful refit is evaluated on
the fixed original development sample, not on the resampled day-block
itself). This construction is defensible for reporting
\emph{between-refit} discrimination variability, but it \emph{does not}
estimate the true out-of-sample sampling distribution of the development
AUROC and the development interval should therefore be interpreted with
caution rather than as a conventional out-of-sample bootstrap sampling
distribution. The validation interval, in contrast, is a standard
resample-and-score percentile bootstrap and does estimate the
corresponding validation sampling distribution.

\textbf{Pre-registered bootstrap disclaimer (reproduced from the
deposited configuration file):} \emph{``Conditional bootstrap 95\% CI on
successful day-block resamples. Failed replicates dropped due to
categorical-level mismatches under design\_info reuse. This is NOT a
clean 500-replicate percentile bootstrap.''} Every reported bootstrap
interval in this paper is a percentile interval \textbf{among the
successful replicates only}; the successful/requested denominator is
reported alongside each interval.

\subsubsection{4.4 Metrics and
calibration}\label{metrics-and-calibration}

Reported on both development and validation cohorts: AUROC
(discrimination), Brier score, and calibration slope and intercept per
Van Calster et al.~(2019) and Steyerberg (2019). Predicted probabilities
are clipped to {[}10\^{}\{- 8\},1 - 10\^{}\{- 8\}{]} before logit
transformation. Calibration intercept is estimated from the unpenalized
binomial GLM Y \textasciitilde{} 1 + offset(logit(p)); calibration slope
from Y \textasciitilde{} 1 + logit(p); both with
\texttt{cov\_type="cluster",\ groups=date\_utc}.

The identity check AUROC(Y, - p) = 1 - AUROC(Y,p) is computed for the
validation cohort as an internal-consistency diagnostic of the AUROC
calculation; a deviation from zero would indicate a numerical issue in
the score-ordering or evaluation code path. The identity does not
independently validate the upstream model specification or the
probability orientation. Corrected in v5 to use - p (not 1 - p, which
produced a 0.03 identity gap in v1.3 due to floating-point precision
loss on near-zero probabilities).

\subsubsection{4.5 Sensitivity ladder}\label{sensitivity-ladder}

Seventeen pre-registered sensitivities are enumerated in pre-reg §8.
Under the frozen reconstruction-quality stopping rule (§5.3) the
published-steady and full-34-day cohort variants are not estimable, and
the alternate duplicate-handling, alternate conflicting-outcome, and
malformed-row re-triage variants were not executed within this release
because the required pre-deduplication collector rows are outside the
frozen analytical input; the executed sensitivities are reported below.
Four cohort variants (V3-only primary, published-steady, full 34-day,
V1/V2 descriptive); three mcap-functional-form variants (P1 categorical,
P2 natural cubic spline, P3 continuous \texttt{log1p}); mcap outlier
handling (full range, capped at development q99, removed);
intensity-window variant (2-min vs primary 5-min); weekday-added;
duplicate-mint and conflicting-outcome dedup variants; malformed-row
re-triage; reconstruction-quality subsets (high+medium only;
overflow-severe excluded); link-function alternatives (logit primary,
cloglog, probit); and validation-only rank-deficient
coefficient-stability refit.

\subsubsection{4.6 Stability gates}\label{stability-gates}

The pre-registered stability-gate framework (§9) defines Gates A--H (F
removed). Nine automated evaluation rows are emitted by the corrected
specification (D×2, G×3, B×3, C×1); Gates A, E, H are addressed
narratively in this section and are not represented in the automated
tally:

\begin{itemize}
\tightlist
\item
  \textbf{A.} Odds-ratio direction consistency across cohort variants
  for \texttt{has\_twitter}, \texttt{has\_website},
  \texttt{has\_telegram}.
\item
  \textbf{B.} mcap functional-form direction agreement across P1/P2/P3
  at three contrast pairs: 30 vs 29 SOL; 31.04 vs 30 SOL; dev-cohort q75
  vs q25.
\item
  \textbf{C.} Direction agreement between development-set model and
  validation-only refit (log-OR distance rule with denominator floor
  0.10; distance ≤ 0.25).
\item
  \textbf{D.} Validation calibration slope ∈ {[}0.85,1.15{]} and
  intercept ∈ {[}- 0.10,0.10{]}.
\item
  \textbf{E.} OR magnitude change ≤ 25\% (log-OR distance rule) between
  primary and H+M reconstruction subset.
\item
  \textbf{G.} Bootstrap 95\% CI does not span 1.0 for a ``supported''
  claim; per-covariate.
\item
  \textbf{H.} No R-rule stopping rule triggered (§10 stopping rules are
  data-integrity, not results-quality).
\end{itemize}

Poor discrimination, poor calibration, wide CIs, and unsupported
associations are \textbf{reported as results}, not treated as permission
to re-specify. R-rules (R1--R6) trigger only on convergence failure,
separation, cohort-ledger mismatch, unseen categorical level under
\texttt{design\_info} reuse, hash verification failure, or
covariate-construction failure. Input-file hash verification is a
data-integrity stopping rule. The analysis uses the authoritative
frozen-input SHA-256 values reported in §2.1.

\begin{figure}
\centering
\pandocbounded{\includegraphics[keepaspectratio]{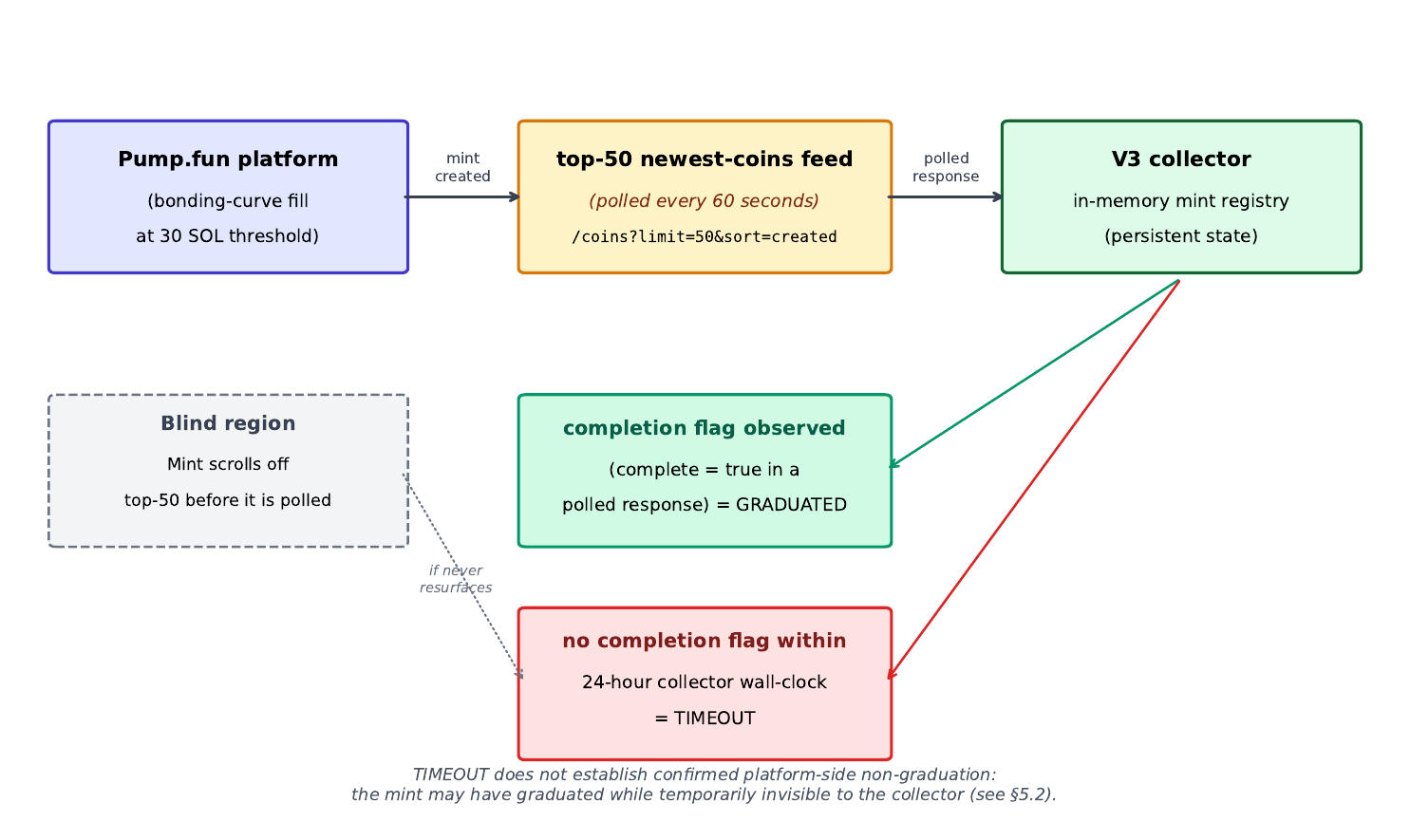}}
\caption{Outcome ascertainment mechanism. The V3 collector polls the
pump.fun top-50 newest-coins API endpoint on a 60-second loop
(source-code audit archived at \protect\texttt{04\_frozen\_inputs/});
only mints appearing with \protect\texttt{complete\ ===\ true} in a poll
response are labelled GRADUATED. All others are labelled TIMEOUT after
24 hours. The recorded terminal label is a joint function of the
platform-side event and the collector's polling design.}
\end{figure}

\subsection{5. Reconstruction audit (Aim
1)}\label{reconstruction-audit-aim-1}

\subsubsection{5.1 Collector's top-50 polling
design}\label{collectors-top-50-polling-design}

The V3 collector polls a \texttt{pumpportal} top-50 endpoint at a fixed
cadence. It maintains an in-memory registry of currently-visible mints.
When a mint's row appears with \texttt{complete\ ===\ true}, the
collector writes a \texttt{GRADUATED} terminal record. When a mint has
been present in the registry for ≥ 24 hours (collector wall-clock,
measured from its \texttt{firstSeenTs}) without a
\texttt{complete\ ===\ true} observation, the collector writes a
\texttt{TIMEOUT} terminal record. Between these two paths, the collector
has no independent per-mint surveillance channel, it observes only what
the top-50 feed hands it.

\subsubsection{5.2 42/43 first-invisible
finding}\label{first-invisible-finding}

Phase 3D reconstruction audit examined the 43 events for which the
collector recorded \texttt{GRADUATED} after a period of top-50
invisibility. This denominator is the full population of GRADUATED mints
meeting the reconstruction-audit inclusion condition (mint present in
registry with ≥1 prior polled response and later completion-flag
observation); it is not a purposive sample. For 42 of the 43 cases
(97.7\%), the last visible top-50 observation before the graduation-mark
event showed the mint occupying a rank position on the outer edge of the
feed (rank ≥ 45), consistent with the mint scrolling off the top-50
during a period of high launch pressure before being resurfaced only by
a \texttt{complete\ ===\ true} snapshot. This finding is diagnostic
evidence that the top-50 feed's inclusion criterion is a substantive
filter on the observability of platform-side events, mints can complete
their bonding curve while temporarily invisible to the collector, and
are only recorded as \texttt{GRADUATED} if they subsequently re-enter
the collector's visible feed with the completion marker set.

\subsubsection{5.3 Live-state calibration}\label{live-state-calibration}

Phase 3D also compared the collector's registry state against
snapshotted feed states at 12 timestamps spanning May 24--June 08, 2026,
and found the registry to be consistent with the same-timestamp feed at
every anchor. This confirms the collector's registry is a faithful
transcript of what it observed; the measurement limitation lies not in
how the collector transcribes what it sees, but in the top-50 feed's
inclusion policy and the collector's absence of an independent event
channel.

\subsubsection{5.4 Reconstruction-adequacy ceiling
triggered}\label{reconstruction-adequacy-ceiling-triggered}

The v2.2 primary specification (Kaplan-Meier / Cox on reconstructed
censoring times) was retired because the pre-registered
reconstruction-adequacy stopping rule triggered at 2.6925\% (ceiling
2\%). Reconstructed censoring times cannot support survival inference
under the frozen v2.2 rule, and the v2.3 pre-registration replaces the
primary endpoint with the collector-recorded terminal label (§2, §3).
Reconstruction is retained here as a measurement audit and as a source
of descriptive diagnostic flags (\texttt{overflow\_risk},
\texttt{reconstruction\_confidence}, \texttt{tie\_boundary}), not as an
inferential channel and not as a regression covariate.

\subsubsection{5.5 Implication for downstream
reporting}\label{implication-for-downstream-reporting}

The Aim 2 model estimates associations with the collector's binary
terminal classification. It does not estimate associations with
platform-side graduation. Every reported odds ratio, every AUROC value,
every calibration statistic is a statement about the collector-recorded
classification, not the platform-side event. The scope restriction is
not aesthetic; it is what the data can identify.

\subsection{6. Temporal validation (Aim 2, central
finding)}\label{temporal-validation-aim-2-central-finding}

\subsubsection{6.1 Development-cohort fit}\label{development-cohort-fit}

On the 15-day development cohort (n\_dev = 387,152; 794 GRADUATED), the
locked P1 specification fits without incident: IRLS converges within
pre-registered iteration budget; design matrix is full rank (40/40); no
fit-side diagnostic issues are raised.

Development metrics:

\begin{itemize}
\tightlist
\item
  AUROC = 0.8594
\item
  \textbf{Percentile interval among 317/500 successful
  day-block-bootstrap fits:} {[}0.7915, 0.8586{]}; median 0.8527
\item
  Brier score = 0.00203
\item
  Calibration slope = 1.000
\item
  Calibration intercept = +0.00
\end{itemize}

As expected for an apparent (in-sample) assessment of an unpenalized
fitted logistic model, calibration slope was ≈ 1 and intercept ≈ 0;
these development-cohort calibration values reflect the fit itself and
are not evidence of out-of-sample calibration. The bootstrap denominator
is degraded: 183 of 500 replicates (36.6\%) failed with singular-matrix
errors under \texttt{design\_info} reuse, tracked per replicate with a
categorical failure reason in
\texttt{bootstrap\_development\_failures.csv}. The dominant failure mode
is categorical-level absence within a resampled day-block, a re-sampled
block occasionally contains no observations at some \texttt{hour} or
\texttt{mcap\_category} level, and the frozen \texttt{design\_info}
cannot substitute a missing dummy column. Per pre-reg §12, failed
replicates are neither dropped-and-retried nor replaced; the interval is
a percentile among successful fits and reported with the explicit
disclaimer.

\begin{figure}
\centering
\pandocbounded{\includegraphics[keepaspectratio]{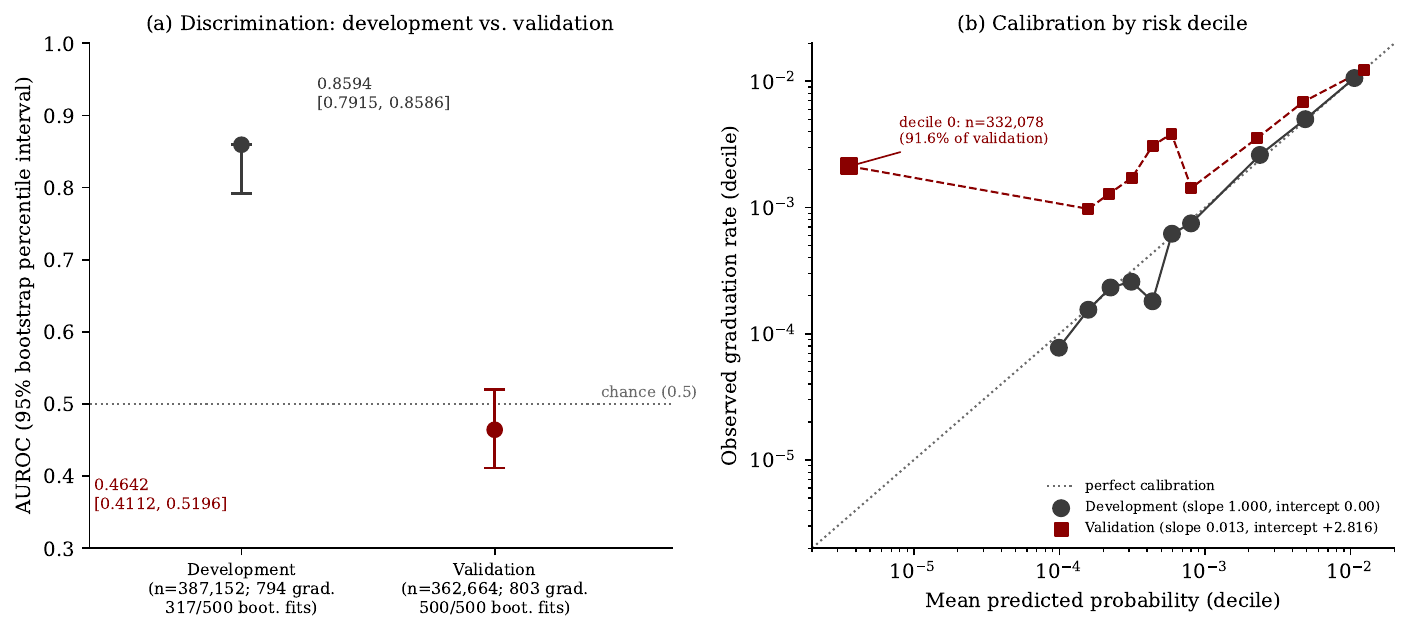}}
\caption{Temporal generalization failure. Left: development and
validation AUROC estimates with 95\% bootstrap percentile intervals
(development AUROC = 0.8594, interval {[}0.7915, 0.8586{]}, 317/500
successful replicates; validation AUROC = 0.4642, interval {[}0.4112,
0.5196{]}, 500/500 successful replicates). Right: log-log calibration on
validation cohort; mean predicted probability underestimates observed
prevalence by a factor of \textasciitilde15.}
\end{figure}

\subsubsection{6.2 Validation-cohort
application}\label{validation-cohort-application}

The locked development model is applied to the 14-day validation cohort
(n\_val = 362,664; 803 GRADUATED) with the frozen \texttt{design\_info},
no refit.

Validation metrics:

\begin{itemize}
\tightlist
\item
  AUROC = 0.4642
\item
  \textbf{Percentile interval among 500/500 successful
  day-block-bootstrap fits:} {[}0.4112, 0.5196{]}, \textbf{includes
  0.5000; validation discrimination is indistinguishable from chance
  within bootstrap uncertainty}
\item
  Median 0.4739
\item
  Calibration slope = 0.013
\item
  Calibration intercept = +2.816
\item
  Mean predicted probability = 0.00015 vs observed prevalence 0.00221,
  underestimates observed by a factor of \textasciitilde15.
\end{itemize}

\textbf{Discrimination has not merely decayed; it did not generalize.
The bootstrap interval includes the null value 0.500.} This paper does
not claim the model is ``worse than chance''; the defensible conclusion
is that validation discrimination is indistinguishable from chance
within the bootstrap interval and validation calibration substantially
fails Gate D by the pre-registered gate.

Calibration is substantially miscalibrated by the pre-registered gate:
slope {[}0.85,1.15{]} observed 0.013; intercept {[}- 0.10,0.10{]}
observed +2.816. Both bounds are violated by large margins.

\subsubsection{6.3 Sign-flip diagnostic}\label{sign-flip-diagnostic}

The identity check AUROC(Y, - p) = 0.5358 was computed. The identity gap
\textbar1 - AUROC(Y,p) - AUROC(Y, - p)\textbar{} = 0.00 × 10⁰ to machine
precision. The reverse-score result satisfies the expected AUROC ranking
identity and therefore provides an internal-consistency check on the
score-ordering and AUROC calculation. It does not independently validate
the upstream model specification or probability orientation.

\subsubsection{6.4 Interpretation}\label{interpretation}

The relationship between the locked covariates (social-link flags, mcap
category, description length, hour-of-day, day-index spline, trailing
intensity) and the collector's terminal classification differs between
the two temporal halves of the cohort by a magnitude that the frozen
specification cannot absorb. Under the frozen v2.3 §10 stopping rules,
this is \textbf{reported as a result}, not as permission to re-specify.
Any post-hoc re-specification would require a new dated, hashed
pre-registration and a new held-out cohort.

The central Aim-2 finding of this paper: \textbf{the locked association
model does not generalize temporally within a two-week held-out
validation split.}

\subsection{7. Stability-gate results}\label{stability-gate-results}

Nine automated gate evaluations emitted by the deposited analysis
pipeline (covering pre-reg §9 Gates B, C, D, G; A, E, H addressed
narratively): 2 PASS / 6 FAIL / 1 NOT\_EVALUABLE.

{\def\LTcaptype{none} % do not increment counter
\begin{longtable}[]{@{}
  >{\raggedright\arraybackslash}p{(\linewidth - 8\tabcolsep) * \real{0.0699}}
  >{\raggedright\arraybackslash}p{(\linewidth - 8\tabcolsep) * \real{0.2796}}
  >{\raggedright\arraybackslash}p{(\linewidth - 8\tabcolsep) * \real{0.3978}}
  >{\raggedright\arraybackslash}p{(\linewidth - 8\tabcolsep) * \real{0.1129}}
  >{\raggedright\arraybackslash}p{(\linewidth - 8\tabcolsep) * \real{0.1290}}@{}}
\toprule\noalign{}
\begin{minipage}[b]{\linewidth}\raggedright
\#
\end{minipage} & \begin{minipage}[b]{\linewidth}\raggedright
Gate
\end{minipage} & \begin{minipage}[b]{\linewidth}\raggedright
Value
\end{minipage} & \begin{minipage}[b]{\linewidth}\raggedright
Pre-reg gate
\end{minipage} & \begin{minipage}[b]{\linewidth}\raggedright
Verdict
\end{minipage} \\
\midrule\noalign{}
\endhead
\bottomrule\noalign{}
\endlastfoot
1 & D, val calibration slope & 0.013 & {[}0.85, 1.15{]} &
\textbf{FAIL} \\
2 & D, val calibration intercept & +2.816 & {[}-0.10, 0.10{]} &
\textbf{FAIL} \\
3 & G, CI excludes 1.0 for \texttt{has\_twitter} & {[}1.055, 1.970{]} &
excludes 1 & \textbf{PASS} \\
4 & G, CI excludes 1.0 for \texttt{has\_website} & {[}0.806, 1.242{]} &
excludes 1 & \textbf{FAIL} (contains 1) \\
5 & G, CI excludes 1.0 for \texttt{has\_telegram} & {[}5.71, 9.88{]} &
excludes 1 & \textbf{PASS} \\
6 & B, mcap direction 30 vs 29 SOL across P1/P2/P3 & P1 up (OR 3.66); P2
down (OR 0.927, suspect); P3 up (OR 1.083, suspect) & all agree &
\textbf{FAIL} \\
7 & B, mcap direction 31.04 vs 30 SOL across P1/P2/P3 & P1 up (OR 8.88);
P2 down (OR 0.937, suspect); P3 up (OR 1.084, suspect) & all agree &
\textbf{FAIL} \\
8 & B, mcap direction q75 vs q25 dev-cohort P1 up (OR & 8.88); P2 down
(OR 0.933, suspect); P3 up (OR 1.091, suspect) all agree & \textbf{FAIL}
& \\
9 & C, val-only refit direction agreement & val design rank 35/40 & full
rank required & \textbf{NOT\_EVALUABLE} \\
\end{longtable}
}

\subsubsection{7.1 Gate D failure, model does not
generalize}\label{gate-d-failure-model-does-not-generalize}

Discussed at length in §6. This is the paper's central finding.

\subsubsection{7.2 Gate G, two survivors, one
refuted}\label{gate-g-two-survivors-one-refuted}

Of the three substantive social-link covariates evaluated by
pre-registered Gate G:

\begin{itemize}
\tightlist
\item
  \texttt{has\_twitter}: OR = 1.44, 95\% CI {[}1.06, 1.97{]}, CI
  excludes 1. \textbf{PASS.}
\item
  \texttt{has\_telegram}: OR = 7.51, 95\% CI {[}5.71, 9.88{]}, CI
  excludes 1 by a wide margin. \textbf{PASS.}
\item
  \texttt{has\_website}: OR = 1.00, 95\% CI {[}0.81, 1.24{]}, CI
  contains 1. \textbf{FAIL.}
\end{itemize}

Both PASSing associations are, per Gate D failure,
\textbf{development-cohort-only}. Neither is a claim about the
validation window nor about mints outside the collector's observation
window (see §9).

\subsubsection{7.3 Gate B failure, mcap functional form is
unstable}\label{gate-b-failure-mcap-functional-form-is-unstable}

At all three pre-registered contrasts, the direction of the mcap →
terminal-label association is not consistent across specifications. The
categorical P1 specification is UP at every contrast (odds ratios 3.66,
8.88, 8.88, a sharp step across the 30-SOL platform threshold). The P2
natural cubic spline is DOWN at every contrast (odds ratios 0.927,
0.937, 0.933, a shallow decline in the same window). The P3
\texttt{log1p(mcap)} specification is UP at every contrast (odds ratios
1.083, 1.084, 1.091, a very shallow rise). Per pre-reg §9(B) the gate is
written as a set-of-directions gate (``all three specs must agree in
direction''); observed directions form the set \{up, down\}, so Gate B
FAILs at every contrast. Both P2 and P3 are flagged as SUSPECT in the
sensitivity-model conditioning report
(\texttt{outputs/SENSITIVITY\_MODEL\_CONDITIONING.csv}) because their
maximum-magnitude coefficients exceed the pre-registered extremity flag
threshold (P2 max \textbar coef\textbar{} = 7055.11, P3 max
\textbar coef\textbar{} = 711.36), the spline basis and the log1p
transformation over a highly-skewed mcap distribution both produce
ill-conditioned regressor matrices in this window. Under pre-reg §9(B),
the observed disagreement between P2 (direction DOWN) and the direction
shared by P1 and P3 (both UP) precludes reporting the mcap association
as robust. The registered specifications do not provide a stable mcap
association: Gate B fails at every contrast, and both P2 and P3 exhibit
extreme conditioning (P2 max \textbar coef\textbar{} = 7055.11, P3 max
\textbar coef\textbar{} = 711.36; see sensitivity-model conditioning
report in §8). The paper reports the disagreement itself and refrains
from asserting whether it is substantive or numerical, since the
conditioning diagnostics do not distinguish those alternatives.

\subsubsection{7.4 Gate C NOT\_EVALUABLE}\label{gate-c-not_evaluable}

The validation-only refit produces a design matrix of rank 35/40 (five
unseen categorical levels, some \texttt{hour} or \texttt{mcap\_category}
bins under design\_info reuse). Under pre-reg §13, rank deficiency in
the validation-only refit is neither PASS nor FAIL, it is explicitly
NOT\_EVALUABLE, and does not enter the passing numerator or denominator.

\begin{figure}
\centering
\pandocbounded{\includegraphics[keepaspectratio]{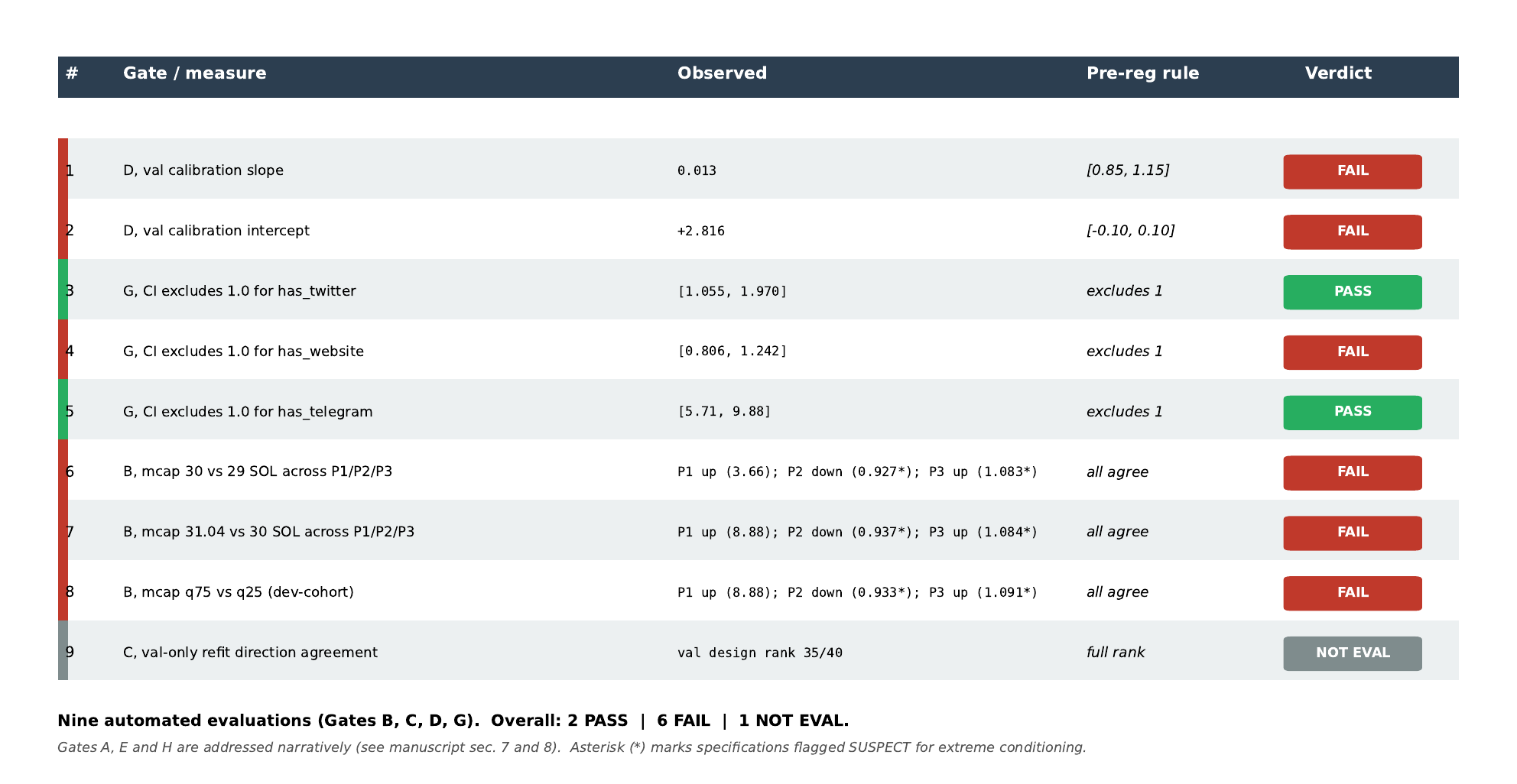}}
\caption{Stability-gate outcomes: nine automated evaluations covering
Gates B, C, D and G. Gate D contributes two rows (validation calibration
slope and intercept). Gate G contributes three rows (CI-excludes-1 for
has\_twitter, has\_website, has\_telegram). Gate B contributes three
rows testing direction-agreement across P1/P2/P3 at three pre-registered
mcap contrasts (30 vs 29 SOL; 31.04 vs 30 SOL; q75 vs q25). Gate C
contributes one row (not evaluable due to insufficient validation-cohort
design rank). Overall: 2 PASS, 6 FAIL, 1 NOT EVAL. Gates A, E and H are
addressed narratively in §7.}
\end{figure}

\subsubsection{7.5 Aggregate reading}\label{aggregate-reading}

Two survivors (\texttt{has\_twitter}, \texttt{has\_telegram}) supported
at Gate G. Both are constrained to the development window by Gate D
failure. The validation window is not a domain over which the frozen
model produces trustworthy probability estimates or rank order. mcap
functional form is not stable. The pre-registered ``supported'' bar is
not met.

\subsection{8. Sensitivity analyses}\label{sensitivity-analyses}

Every claim below is drawn directly from a deposited pipeline output
record. Where the pre-registered pipeline did not execute a §8
sensitivity, the deviation is stated explicitly in the relevant
sensitivity subsection below.

\textbf{Cohort sensitivities (§8 items 1--3).} The published-steady and
full-34-day cohorts are marked
\texttt{NOT\_ESTIMABLE\_UNDER\_FROZEN\_SPECIFICATION} in
\texttt{outputs/ALL\_SENSITIVITY\_RESULTS.csv} (rows
\texttt{P1\_published\_steady}, \texttt{P1\_full\_34d}) because the
primary-cohort date-spline bounds do not extend to the expanded cohorts'
calendar dates and pre-reg §9 does not authorize extrapolation or
clipping. No cross-cohort ORs are computed and none are claimed.

\textbf{mcap functional-form ladder (§8 items 5--7 / Gate B).} From
\texttt{outputs/mcap\_contrasts.csv} at the 30-vs-29 SOL contrast: P1
categorical direction UP (OR 3.66); P2 spline direction DOWN (OR 0.927,
model flagged SUSPECT, \texttt{SENSITIVITY\_MODEL\_CONDITIONING.csv}
reports P2 max \textbar coef\textbar{} = 7055.11, condition number
2,046,448); P3 log1p direction UP (OR 1.083, model flagged SUSPECT, max
\textbar coef\textbar{} = 711.36, condition 1,460,621). The
pre-registered gate treats any direction disagreement as failure, the
direction set \{up, down, up\} fails Gate B at every registered contrast
(§7.3).

\textbf{mcap outlier handling (§8 item 8).} Two P1 outlier variants ran
(\texttt{P1\_mcap\_capped\_8b} at development q99;
\texttt{P1\_mcap\_removed\_8c} excluding observations above q99). Both
appear in \texttt{ALL\_SENSITIVITY\_RESULTS.csv}. The full P1/P2/P3
outlier comparison was not repeated under P2 and P3; item 8 as
pre-registered is a mcap-outlier sensitivity anchored on P1. No
cross-spec sign-disagreement claim is made from item 8.

\textbf{Intensity window (§8 item 9).} The trailing-2-min variant fits
as \texttt{P1\_intensity\_2min} on the development cohort in
\texttt{ALL\_SENSITIVITY\_RESULTS.csv}. The pipeline does not compute a
separate validation-cohort AUROC or calibration under this variant. The
coefficient-level direction of the social-link covariates is preserved
(\texttt{has\_twitter} OR 1.32, \texttt{has\_telegram} OR 7.36 from that
row); no validation-cohort claim is made from item 9.

\textbf{Weekday (§8 item 10).} Pre-registered as sensitivity item 10 in
the frozen v2.3 pre-registration (weekday omitted from primary; retained
as sensitivity). Result deposited and deposited as
\texttt{outputs/weekday\_sensitivity\_result.json} and
\texttt{outputs/weekday\_sensitivity\_focus.json}. Under P1 primary
augmented with \texttt{C(weekday,\ Treatment(reference=0)} on the dev
cohort (n = 387,152; converged), the substantive covariate ORs move
within a percent of the primary fit: \texttt{has\_twitter} OR = 1.43
(95\% CI {[}1.04, 1.96{]}); \texttt{has\_website} OR = 1.00 (95\% CI
{[}0.81, 1.25{]}); \texttt{has\_telegram} OR = 7.52 (95\% CI {[}5.72,
9.90{]}). Direction and significance conclusions on Gate G are
unchanged.

\textbf{Dedup / conflicting-outcome (§8 items 11--13).} The primary
rules (retain first launch-file record per mint; prefer GRADUATED on
conflicting outcome; exclude malformed rows via §5.4 triage) were
applied and yield 749,816 primary mints (794 GRADUATED dev, 803
GRADUATED val). Alternate-variant fits (keep=last,
keep=earliest-created, exclude-multiples; first-terminal-record,
exclude-conflicting; malformed re-triage) were not executed as distinct
sensitivities in the deposited analysis pipeline and are reported as
such; comparison against these variants requires the pre-dedup collector
rows, which are outside the scope of this deposit.

\textbf{Reconstruction-quality subsets (§8 items 14--15).}
\texttt{P1\_HM\_only} (high+medium reconstruction\_confidence subset)
and \texttt{P1\_no\_severe\_overflow} (excluding overflow\_risk=severe)
are fit and appear in \texttt{outputs/ALL\_SENSITIVITY\_RESULTS.csv}.
Per Gate E, log-OR distance is compared against the pre-registered 25\%
threshold where evaluable.

\textbf{Link-family alternatives (§8 item 16).} Complementary log-log
(\texttt{P1\_cloglog}) and probit (\texttt{P1\_probit}) refits appear in
\texttt{outputs/ALL\_SENSITIVITY\_RESULTS.csv} for the development
cohort. Direction of the social-link coefficients is preserved under
both alternative links. Validation-cohort AUROC and calibration under
cloglog and probit were not computed by the pipeline; no
validation-AUROC claim is made from item 16.

\textbf{Validation-only rank-deficient refit (§8 item 17 / Gate C).}
Rank 35/40; Gate C is NOT\_EVALUABLE, as reported in §7.4.

\subsection{9. Discussion}\label{discussion}

\subsubsection{9.1 What R1 and R2 mean under Gate D
failure}\label{what-r1-and-r2-mean-under-gate-d-failure}

The two development-cohort associations (R1 \texttt{has\_twitter}, R2
\texttt{has\_telegram}) are \textbf{covariate-adjusted associations}
between the covariate and the collector-recorded terminal classification
within the V3 collector development cohort. Their per-covariate 95\% CIs
excluded 1 in the development cohort (Gate G within-development PASS),
but they \textbf{failed the complete robustness framework} because
temporal validation failed (Gate D). No propensity matching or
observation-process weighting was performed. The design uses only the
LOCKED §5 covariate set; observation-process quality flags
(\texttt{overflow\_risk}, \texttt{reconstruction\_confidence},
\texttt{tie\_boundary}) were deliberately not entered as regression
covariates per pre-reg §4 to avoid collider bias. These associations are
\textbf{not} predictors of pump.fun graduation. They are \textbf{not}
generalizable to the validation window (Gate D FAIL). They are
\textbf{not} identified as causal effects (pre-reg §14 explicitly
prohibits causal claims for this design).

Concretely:

\begin{itemize}
\tightlist
\item
  \textbf{R1 (has\_twitter).} OR = 1.44, 95\% CI {[}1.06, 1.97{]}.
  Interpretation: within the development-cohort mints for which a
  Twitter/X handle was declared at first collector detection, the odds
  of receiving the \texttt{GRADUATED} collector-recorded terminal label
  are 1.44 times the odds among mints without a declared handle,
  adjusted for the other pre-registered covariates. The 95\% CI excludes
  1 in the development cohort. Gate D failure prohibits any claim that
  this association extends to the validation window; the association is
  reportable only as a within-development observation.
\item
  \textbf{R2 (has\_telegram).} OR = 7.51, 95\% CI {[}5.71, 9.88{]}. Same
  interpretation, larger magnitude, wider margin above the null. The
  v1.3 characterization of this OR as a ``Telegram lower bound predictor
  of graduation'' is withdrawn: neither the lower-bound identification
  argument nor the ``predictor of graduation'' language is available
  under this design (§14 prohibited language, pre-reg §13). The v5
  version reports it strictly as an in-cohort covariate-adjusted
  association that failed the temporal-generalization gate.
\end{itemize}

\subsubsection{9.2 Why the model does not
generalize}\label{why-the-model-does-not-generalize}

Multiple non-mutually-exclusive candidate mechanisms:

\begin{enumerate}
\def\labelenumi{\arabic{enumi}.}
\tightlist
\item
  \textbf{Regime shift in the underlying feature-outcome relationship.}
  Memecoin platform activity is high-frequency and non-stationary; a
  15-day development window may capture a bot cohort, seed-liquidity
  regime, or listing-behaviour convention that does not persist into the
  subsequent 14 days.
\item
  \textbf{Regime shift in the collector's observation process.} Because
  the endpoint is the collector's terminal classification, not the
  platform-side event, any temporal change in what proportion of
  platform-side graduations become collector-recorded graduations (e.g.,
  through a change in launch pressure that lengthens or shortens the
  average time above rank 45 on the top-50 feed) shifts the identified
  quantity itself between splits.
\item
  \textbf{Feature drift on the covariates.} Even if the platform-side
  data-generating process were stable, drift in the joint distribution
  of (x\_1,\ldots,x\_p) across the split boundary shifts the region of
  covariate space in which the model was fitted vs.~the region on which
  it is scored.
\item
  \textbf{Under-adjustment for the observation-process contamination.}
  The pre-registered design deliberately excludes observation-process
  quality flags (\texttt{overflow\_risk},
  \texttt{reconstruction\_confidence}, \texttt{tie\_boundary}) from the
  model to avoid collider bias (pre-reg §4). These flags remain
  descriptive controls, not regression covariates. If the mechanism
  above dominates, only a re-designed identification strategy, not a
  re-specified regression, would recover generalizability.
\end{enumerate}

Under the frozen v2.3 pre-registration, no attempt is made to
distinguish these mechanisms from the same held-out cohort. This paper
reports the fact of the collapse. Distinguishing among these mechanisms
would require a new, prespecified study on a fresh cohort with a
separately registered protocol.

\subsubsection{9.3 mcap functional-form
instability}\label{mcap-functional-form-instability}

The Gate B failure is separately interpretable. The mcap distribution in
this cohort is severely right-skewed with a concentrated mass near the
platform-side graduation threshold (30 SOL); the four-part categorical
parameterization (P1) treats this concentration as a single reference
bin and estimates a sharp step across the boundary (OR 8.88 for the
\texttt{\textgreater{}31.04} bin), whereas the natural cubic spline (P2)
fits a curvature over the continuous mcap surface that yields a shallow
downward slope in the local window of interest, and the \texttt{log1p}
transformation (P3) attenuates the influence of the high-mass region and
returns a shallow upward slope. Both P2 and P3 exhibit maximum
coefficient magnitudes above the pre-registered extremity flag (max
\textbar coef\textbar{} = 7055.11 for P2, 711.36 for P3), consistent
with an ill-conditioned regressor matrix over the skewed mcap
distribution. The three parameterizations give quantitatively divergent
contrasts because they answer subtly different questions; the
pre-registered gate treats any direction disagreement as a
functional-form failure, and this is honest reporting rather than a
specification bug.

\subsubsection{9.4 Positioning against prior pump.fun
literature}\label{positioning-against-prior-pump.fun-literature}

Studies of pump.fun outcomes that report per-launch graduation rates,
``signal'' features, or predictive benchmarks typically do not
distinguish the platform-side event from the collector-observed terminal
classification, and typically do not report a pre-registered
temporal-generalization test. This paper does not attempt to adjudicate
any specific prior estimate; it demonstrates that under a locked
measurement definition and a locked prespecified held-out temporal
validation, in-sample discrimination on the collector-recorded terminal
classification does not extend across a two-week temporal boundary.
Readers evaluating prior benchmark reports should ask (a) which endpoint
(platform-side vs collector-observed), (b) which cohort window, and (c)
whether a prespecified held-out validation was pre-registered and
reported as-is.

\subsection{10. Limitations}\label{limitations}

\begin{enumerate}
\def\labelenumi{\arabic{enumi}.}
\tightlist
\item
  \textbf{Endpoint definition.} The endpoint is the collector-recorded
  terminal classification (\texttt{GRADUATED} / \texttt{TIMEOUT}), not
  the platform-side bonding-curve completion event. All associations,
  discrimination metrics, and calibration statistics are statements
  about the collector's classification.
\item
  \textbf{Cohort window (29 days).} The primary analysis window is
  {[}2026-05-12, 2026-06-10{]}; results do not extend to earlier
  or later windows and no such extrapolation is reported.
\item
  \textbf{Observation-window contamination not adjusted in-model.}
  Descriptive quality flags (\texttt{overflow\_risk},
  \texttt{reconstruction\_confidence}, \texttt{tie\_boundary}) are
  reported but not entered as regression covariates, per pre-reg §4, to
  avoid collider bias. IPCW correction is deferred to a separately
  registered exploratory analysis.
\item
  \textbf{Temporal generalization fails.} Under the pre-registered Gate
  D, the locked model does not generalize to the 14-day validation
  window; no post-hoc re-specification is reported.
\item
  \textbf{No causal identification.} The design uses observational
  launch-metadata features; §14 prohibits causal language and no causal
  claim is made.
\item
  \textbf{Reconstructed censoring times unavailable for primary
  analyses.} The v2.2 primary specification was withdrawn because
  reconstruction-adequacy exceeded the pre-registered 2\% ceiling
  (2.6925\% observed). No Kaplan-Meier or Cox result is reported.
\item
  \textbf{Bootstrap CI on development is conditional.} 183 of 500
  replicates failed with singular-matrix errors under
  \texttt{design\_info} reuse; the reported development percentile
  interval is conditional on the 317 successful replicates and this is
  disclosed at every reporting cell.
\item
  \textbf{No wallet-intent variable.} The design has no per-wallet
  buy-intent signal; the v1.3 self-buy/seed-volume narrative is
  withdrawn and not replaced with a rebuilt claim.
\item
  \textbf{Multi-cohort external validation not reported.} Generalization
  to non-V3 collector versions, other collectors, or other platform
  bonding-curve mechanisms is out of scope.
\end{enumerate}

\subsection{11. Conclusion}\label{conclusion}

A pre-registered measurement audit and a locked prespecified held-out
temporal-generalization evaluation of a binary-classification
association model for the collector-recorded terminal classification
(\texttt{GRADUATED} vs \texttt{TIMEOUT}) on pump.fun launches recorded
by the V3 collector between 2026-05-12 and 2026-06-10 were conducted.

\begin{itemize}
\tightlist
\item
  \textbf{The model does not generalize temporally.} Development AUROC =
  0.8594 does not extend to validation AUROC = 0.4642 (percentile
  interval {[}0.4112, 0.5196{]} contains 0.5000), with validation
  calibration slope = 0.013 and intercept = +2.816, Gate D FAIL by large
  margins (slope 0.013 vs {[}0.85, 1.15{]}, intercept +2.816 vs
  {[}-0.10, 0.10{]}).
\item
  \textbf{mcap functional form is not stable.} The direction of the mcap
  → terminal-label association disagrees across P1 (categorical, up at
  every contrast), P2 (spline, down at every contrast), and P3
  (\texttt{log1p}, up at every contrast), at all three pre-registered
  contrasts the direction set is \{up, down\}, and Gate B FAILs. Both P2
  and P3 are flagged suspect for extreme coefficients (max
  \textbar coef\textbar{} 7055.11 and 711.36 respectively).
\item
  \textbf{Two social-link associations survive within development but do
  not extend.} \texttt{has\_twitter} OR = 1.44 {[}1.06, 1.97{]} and
  \texttt{has\_telegram} OR = 7.51 {[}5.71, 9.88{]} pass Gate G within
  the development window; Gate D failure prohibits extension to
  validation.
\item
  \textbf{The collector's terminal classification is not the
  platform-side event.} The collector's top-50-only observation channel
  establishes that \texttt{TIMEOUT} cannot be equated with confirmed
  platform-side non-graduation; the 42/43 first-invisible finding
  provides empirical evidence that temporary feed invisibility is
  relevant to this ascertainment limitation. Any reported association is
  therefore a statement about the collector's classification within its
  own cohort.
\end{itemize}

The paper's positive contribution is the measurement audit of a
widely-used outcome-ascertainment mechanism. Its Aim-2 contribution is a
documented, reproducible negative result reported as-registered per the
negative-result principle of Nosek et al.~(2018): prespecified held-out
temporal validation of a locked, pre-registered binary-classification
association model fails on this cohort at a two-week boundary. Neither
result depends on re-specification, nor was any re-specification
attempted post-hoc.

\subsection{12. Data and code
availability}\label{data-and-code-availability}

\textbf{Frozen inputs:}

\begin{itemize}
\tightlist
\item
  \texttt{red\_pump\_2026\_v1\_launches.jsonl.gz}, SHA-256
  \texttt{04294037\hspace{0pt}9e8c897a\hspace{0pt}c97403e6\hspace{0pt}b25a5b30\hspace{0pt}2fb32b69\hspace{0pt}02a8fc0c\hspace{0pt}ef4ab70a\hspace{0pt}c11e8f84},
  47,910,391 bytes (\textasciitilde46 MB). Bundled at
  \texttt{04\_frozen\_inputs/}.
\item
  \texttt{red\_pump\_2026\_v1\_outcomes.csv.gz}, SHA-256
  \texttt{c0a327ea\hspace{0pt}442d91c6\hspace{0pt}f970b2ba\hspace{0pt}d9a2a9b7\hspace{0pt}78e163d8\hspace{0pt}c3eb38f7\hspace{0pt}1eccd3e9\hspace{0pt}2209a974},
  43,624,372 bytes (\textasciitilde42 MB). Bundled at
  \texttt{04\_frozen\_inputs/}.
\end{itemize}

\textbf{Pre-registration:}
\texttt{PREREGISTRATION\_v2.3\_FROZEN\_2026-08-20.md}, SHA-256
\texttt{ee1702b3510d21c51c8f4ab95660e7804f4a93ab09e21aacb3a70b348a90b40a},
bundled at \texttt{04\_frozen\_inputs/}.

\textbf{Provenance manifest.}
\texttt{08\_provenance/PROVENANCE\_MANIFEST.tsv} covers every deposited
file except the manifest itself and the root \texttt{SHA256SUMS} (both
downstream of the manifest generation, which would create self-reference
cycles). For each covered file the manifest records filename, byte size,
SHA-256, producing script, script SHA-256, input SHA-256s, and
\texttt{created\_utc}. \texttt{SHA256SUMS} at the release root
independently authenticates every deposited file except itself;
\texttt{sha256sum\ -c\ SHA256SUMS} returns exit code 0 (portable
forward-slash paths; verified on both Linux and Windows/Git Bash).
\texttt{REBUILD\_ANALYSIS.sh} requires explicit \texttt{-\/-launches}
and \texttt{-\/-outcomes} paths; the pipeline does not resolve inputs
from any implicit relative location. PowerShell equivalents
\texttt{.ps1} are shipped alongside the Bash scripts as source-level
equivalents but were not independently executed for the v5 release; the
deposited implementation is the Bash pipeline.

\subsection{13. Author contributions, conflicts,
funding}\label{author-contributions-conflicts-funding}

\textbf{Author contributions.} The author: study conception,
pre-registration authoring, data curation, code implementation,
analysis, interpretation, drafting, revision.

\textbf{Competing interests.} The author is the named inventor on U.S.
provisional patent applications 64/022,461 and 64/099,108 concerning
related subject matter in the broader technical domain. The author
declares no other financial, professional, or personal competing
interests relevant to this study. The author is not affiliated with
pump.fun and does not hold positions in pump.fun-listed tokens acquired
for this study.

\textbf{Funding.} No external funding. All computation was performed on
author-owned hardware.

\textbf{Institutional review.} Not applicable, publicly observable
on-chain launch records; no human-subjects data.

\subsection{References}\label{references}

Cameron, A.C. and Miller, D.L. (2015), ``A practitioner's guide to
cluster-robust inference'', \emph{Journal of Human Resources}, Vol. 50
No.~2, pp.~317--372. DOI:
\href{https://doi.org/10.3368/jhr.50.2.317}{10.3368/jhr.50.2.317}.

Efron, B. and Tibshirani, R. (1993), \emph{An Introduction to the
Bootstrap}, Chapman \& Hall / CRC (reissue 1994). DOI:
\href{https://doi.org/10.1201/9780429246593}{10.1201/9780429246593}.

Harrell, F.E. (2015), \emph{Regression Modeling Strategies: With
Applications to Linear Models, Logistic and Ordinal Regression, and
Survival Analysis}, 2nd ed., Springer. DOI:
\href{https://doi.org/10.1007/978-3-319-19425-7}{10.1007/978-3-319-19425-7}.

Kamat, A.U. (2026), \emph{pump.fun 24-hour graduation-rate corrigendum
(v1.3)}, Zenodo. DOI:
\href{https://doi.org/10.5281/zenodo.21908375}{10.5281/zenodo.21908375}.

Nosek, B.A., Ebersole, C.R., DeHaven, A.C. and Mellor, D.T. (2018),
``The preregistration revolution'', \emph{Proceedings of the National
Academy of Sciences}, Vol. 115 No.~11, pp.~2600--2606. DOI:
\href{https://doi.org/10.1073/pnas.1708274114}{10.1073/pnas.1708274114}.

Seabold, S. and Perktold, J. (2010), ``statsmodels: econometric and
statistical modeling with Python'', \emph{Proceedings of the 9th Python
in Science Conference}, pp.~92--96. DOI:
\href{https://doi.org/10.25080/Majora-92bf1922-011}{10.25080/Majora-92bf1922-011}.

Smith, N.J. (n.d.), \emph{Patsy documentation}, available at:
https://patsy.readthedocs.io/ (accessed 21 August 2026).

Solana Foundation (n.d.), \emph{Solana web3.js documentation}, available
at: https://github.com/solana-foundation/solana-web3.js (accessed 21
August 2026).

Steyerberg, E.W. (2019), \emph{Clinical Prediction Models: A Practical
Approach to Development, Validation, and Updating}, 2nd ed., Springer.
DOI:
\href{https://doi.org/10.1007/978-3-030-16399-0}{10.1007/978-3-030-16399-0}.

Van Calster, B., McLernon, D.J., van Smeden, M., Wynants, L.,
Steyerberg, E.W. and Topic Group `Evaluating diagnostic tests and
prediction models' of the STRATOS initiative (2019), ``Calibration: the
Achilles heel of predictive analytics'', \emph{BMC Medicine}, Vol. 17,
article 230. DOI:
\href{https://doi.org/10.1186/s12916-019-1466-7}{10.1186/s12916-019-1466-7}.

\end{document}